\documentclass[fleqn,10pt]{wlscirep}
\usepackage[utf8]{inputenc}
\usepackage[T1]{fontenc}
\usepackage{array}
\usepackage{lscape}

\title{30-min Decayless Kink Oscillations in a Very Long Bundle of Solar Coronal Plasma Loops} 
\author[1]{Sihui Zhong}
\author[1,2,*]{Valery M. Nakariakov}
\author[3]{Yuhu Miao}
\author[4,5]{Libo Fu}
\author[4,5,*]{Ding Yuan}
\affil[1]{Centre for Fusion, Space and Astrophysics, Physics Department, University of Warwick, Coventry CV4 7AL, UK}
\affil[2]{Centro de Investigacion en Astronom\'ia, Universidad Bernardo O'Higgins, Avenida Viel 1497, Santiago, Chile}
\affil[3]{School of Information and Communication, Shenzhen Institute of Information Technology, Shenzhen 518172, China}
\affil[4]{Institute of Space Science and Applied Technology, Harbin Institute of Technology, Shenzhen, Guandong 518055, China}
\affil[5]{Shenzhen Key Laboratory of Numerical Prediction for Space Storm, Harbin Institute of Technology, Shenzhen, Guandong 518055, China}
\affil[*]{Corresponding authors: v.nakariakov@warwick.ac.uk, yuanding@hit.edu.cn}

\begin{abstract}

The energy balance in the corona of the Sun is the key to the long-standing coronal heating dilemma, which could be potentially revealed by observational studies of decayless kink oscillations of coronal plasma loops.
A bundle of very long off-limb coronal loops with the length of $736\pm80$\,Mm and a lifetime of about 2 days are found to exhibit decayless kink oscillations. The oscillations were observed for several hours. The oscillation amplitude was measured at 0.3--0.5\,Mm, and the period at 28--33\,min.
The existence of 30-min periodicity of decayless kink oscillations indicates that the mechanism compensating the wave damping is still valid in such a massive plasma structure. 
It provides important evidence for the non-resonant origin of decayless kink oscillations with 2--6~min periods, i.e., the lack of their link with the leakage of photospheric and chromospheric oscillations into the corona and the likely role of the broadband energy sources.
Magnetohydrodynamic seismology based on the reported detection of the kink oscillation, with the assistance of the differential emission measure analysis and a background coronal model provides us with a comprehensive set of plasma and magnetic field diagnostics, which is of interest as input parameters of space weather models. 

\end{abstract}
\begin{document}

\flushbottom
\maketitle
%
%
\thispagestyle{empty}


\section*{Introduction}
\label{sec:intro}


The outermost part of the solar atmosphere, the corona, is a layer of fully-ionised plasma with a temperature of about 200-500 times greater than the visible surface of the Sun, which is closer to the nuclear fusion energy source of the Sun. This is a century-long challenge in physics, the solar coronal heating problem which requires to reveal how the corona is maintained at millions of degrees. 
Two tentative heating mechanisms are related to omnipresent magnetohydrodynamic (MHD) waves \cite{1947MNRAS.107..211A,2020SSRv..216..140V} and an enormous sum of discrete, small-scale bursts, i.e., nanoflares \cite{1988ApJ...330..474P}.
The quantitative observational examination of different mechanisms needs detailed physical conditions in the corona, particularly, the magnetic field.
Similar to seismic inversions in geophysics, coronal seismology, the study of coronal MHD waves, allows remote plasma diagnostics \cite{1984ApJ...279..857R,2020ARA&A..58..441N}. 

Among all types of MHD waves which can be hosted by plasma non-uniformities of the corona, e.g., coronal loops, kink oscillations are subject to intensive investigation, as they are spatially and temporally resolvable by imaging instruments since the very first launch of high-resolution spacecraft (TRACE, Transition Region And Coronal Explorer) {\cite{1999Sci...285..862N,1999ApJ...520..880A,2002ApJ...576L.153O}}.
Kink oscillations of coronal loops exhibit periodic transverse displacement, and appear in two distinct regimes. One regime has large amplitude but decays rapidly, hence named decaying kink oscillations; another regime is low-amplitude and remains undamped, hence decayless. 
In observations of both regimes, linear scaling of the loop length and oscillation period is established, with the range of loop lengths of 100--789\,Mm, and oscillation period of 1-28\,min in the decaying regime \cite{2002SoPh..206...99A,2003ApJ...598.1375A,2004SoPh..223...77V,2015A&A...577A...4Z,2016A&A...585A.137G,2019ApJS..241...31N}, and of 11-560\,Mm and 11\,s-10.33\,min, respectively, in the decayless regime \cite{2015A&A...583A.136A,2021A&A...652L...3M,2023ApJ...944....8L}. This empirically determined linear scaling reveals the standing wave nature of this phenomenon.
{It has also been proposed that the persistent loop oscillations could heat the corona and power the solar wind \cite{2011Natur.475..477M,2012ApJ...759..144T}.}
Seismological inversions of kink oscillations of loops give us estimations of the Alfv\'en speed, density scale height, magnetic field strength, etc \cite{2021SSRv..217...73N}.
For seismological applications, decayless kink oscillations are an attractive tool, due to their great potential as a routine probe, especially before flares and eruptions. This is because the occurrence rate of decaying kink oscillations is low as they are sparked by low-occurrence impulsive energy releases, whereas decayless kink oscillations are ubiquitous in quiescent active regions.
Moreover, the physical properties estimated via seismology can be used to attempt particular mechanisms predicted by theories. For example, the resonant absorption mechanism has been tested with the empirical density contrast, the ratio of internal and external density, and the thickness of the inhomogeneous layer of the loop, demonstrating its consistency with observational properties of the oscillations \cite{2003ApJ...598.1375A} (see, however, \cite{2008ApJ...676L..77A} for an alternative view). 
On the other hand, one cannot rule out other damping mechanisms, e.g., the leakage of the oscillations into the corona by the excitation of outwardly propagating fast magnetoacoustic waves, {see, e.g.\cite{2007A&A...462.1127S,2010ApJ...714..170S, 2011ApJ...728...87S}, caused by 3D effects}.

On the other hand, in the context of the coronal energy balance, of interest is the undamped nature of decayless kink oscillations, opposite to {another} type of kink oscillations that damp quickly. Importantly, the duration of this phenomenon can be several tens of oscillation cycles, without a significant variation in amplitude and period \cite{2022MNRAS.513.1834Z}. 
The aforementioned resonant absorption \cite{2002A&A...394L..39G,2007A&A...463..333A,2008A&A...484..851G,2020SSRv..216..140V} {as well as the wave leakage \cite{2007A&A...462.1127S, 2010ApJ...714..170S}}, work in the linear regime {\cite{2019ApJS..241...31N}}, so it could play a role in low-amplitude decayless oscillations.  
Independently of the specific mechanism for the rapid damping of impulsively excited kink oscillations, the physical mechanism to compensate the energy losses in the decayless regime may be a clue to unravel the coronal heating problem. In particular, it may indicate the nature of the energy supply into the corona. Yet the driving mechanism of decayless kink oscillations is unanswered. Currently, balancing candidates are the interaction between loop footpoints and the quasi-steady flows \cite{2016A&A...591L...5N,2020ApJ...897L..35K,2021ApJ...908L...7K} or random flows \cite{2020A&A...633L...8A,2021MNRAS.501.3017R}. In both models, the periods of the excited kink oscillations are set by the loop itself, rather by the driving periods. Alternatively, decayless kink oscillations could be seen as apparent periodic brightness associated with Kelvin-Helmholtz Instability (KHI) \cite{2016ApJ...830L..22A} and interference fringe in a corona arcade \cite{2014ApJ...784..103H}.
Each proposed mechanism has distinct signatures in the energy source, transportation, and conversion. Especially, the energy spectrum of the driver could probably be uncovered by the existence of observational cases of extreme periodicity, i.e., the lower and upper limit of time scales.

Here, we present an observational detection of a decayless kink oscillation of a bundle of very long loops, with oscillation period of around 30\,min, which makes it the longest periodicity of a decayless kink oscillation ever reported. Additionally, we perform a comprehensive seismological inversion to diagnose the loop properties including density, temperature, kink and Alfv\'en speeds, magnetic field strength, etc.
This case study provides important evidence to support the non-resonantly-driven nature of decayless kink oscillations.

\section*{Results}
\label{sec:results}

The Atmospheric Imaging Assembly (AIA) \cite{2012SoPh..275...17L} on board the Solar Dynamics Observatory \cite{2012SoPh..275....3P}, captured a set of long-living very long loops near the North-West limb from 2022-11-08T12:00 to 2022-11-10T20:30\,UT in several passbands including 94\,\AA, 131\,\AA, 171\,\AA, and 193\,\AA. In the 171\,\AA\ channel, the loops had the highest contrast with the background. 
During this two-day period, the loops of interest are dynamic, and most of the time the loop part off the limb is well visible. However, the whole loop connectivity is obscured, as the loop footpoints could be only seen in a certain time.
The geometry of the loops is indicated in Figure \ref{fig:im}(a--b).
The west-southern footpoint is located in AR 13135, while the east-northern one is in AR 13137, see the plus symbols in Figure \ref{fig:im}(b). Hence this is a trans-active region coronal loop.
The length of the loop shown in Figure \ref{fig:im}(b) was estimated by the distance between two footpoints, assuming that the loop is of a semi-circular shape. The Helioprojective longitudes and latitudes of the footpoints are $[43.63^{\circ}, 47.22^{\circ}],[87.36^{\circ}, 27.17^{\circ]}$, respectively. It gives us the chord length of 469\,Mm, the loop length of 736\,Mm, and the major radius of 234\,Mm. The sampling error is $5^{\circ}$ in both two dimensions, then the uncertainty of loop length is about 80\,Mm.

At around 2022-11-10T14:30 UT, kink oscillations of the northern loop leg can be clearly observed by the naked eye, lasting longer than 2 hours. Additional kink oscillations were revealed with the help of the motion magnification technique (see Methods). The detection ceases when the loops fade away because of the change of observation conditions. The oscillations do not show any systematic damping during its detection. 
During the oscillation, there were no flares and eruptions. The nearest in time solar flare of a GOES B6.7 class occurred in AR 13135 from 11-09T22:50 to 11-10T00:17\,UT, and hence could not be linked to the oscillation. Hence, the detected oscillation is of the decayless kind.


\subsection*{Oscillation properties}
\label{sec:osci}
\begin{figure}[ht]
\centering
\includegraphics[width=\linewidth]{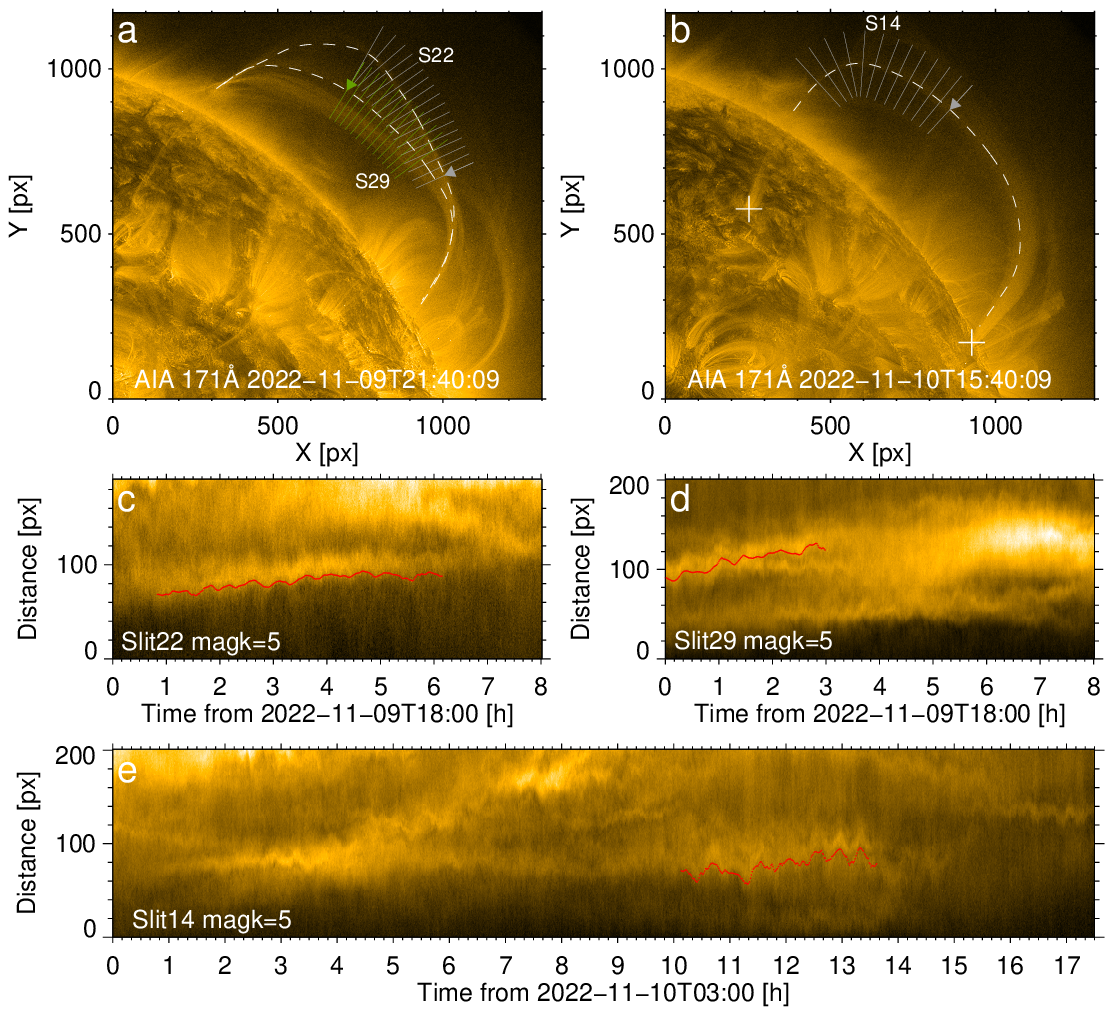}
\caption{The observed decayless kink oscillations. Panels (a--b): EUV 171\,\AA\ images of the field of interest at 2022-11-09T21:40~UT and 2022-11-10T15:40~UT. The oscillating loops are tracked by the dashed curves. The slits perpendicular to the loop segments are used to make time--distance maps (d--e) that reveals the oscillatory patterns. The arrows indicate the direction of distance along the slits. The transverse oscillations evident in time-distance maps are magnified by a factor of 5. The displacement of the loop axis is marked by the red curves with a 10 pixel offset in the vertical direction.}
\label{fig:im}
\end{figure}

During 2022-11-09T18:00 to 2022-11-10T20:30, the loops which form the bundle of interest exhibit three events of decayless kink oscillations. The oscillations are seen in two time intervals, at around 2022-11-09T21:40 UT (see Figure~\ref{fig:im}a) and 2022-11-10T15:40 UT (see Figure~\ref{fig:im}b). Details of the transverse oscillatory patterns were determined with the use of time--distance maps constructed for 1D slits across particular loop segments during those time intervals.  
Oscillations detected in different segments of the loop appear to be in phase with each other. 
In the figure, we demonstrate time--distance maps made using a certain slit across each loop, the maps made for the neighbouring slits are similar.
In some segments of the loops and some time intervals, the oscillations are difficult to detect, which is attributed to less favourable observation conditions, e.g., overlapping of other strands, blurring by background noise, etc. 

Panels~\ref{fig:im}(c--d) depict decayless kink oscillations in the time interval from 2022-11-09T18:00 to 2022-10-10T02:00. In panel~(c), the middle oscillatory pattern (marked by the red curve) is from the upper loop indicated by the outer white curve in panel~(a). 
This oscillation lasts for more than 5\,h, with the oscillation period longer than 20\,min, and the average transverse displacement amplitude of 0.26\,Mm.
In panel~(d), oscillations of two different loops are clearly visible, in the middle loop tracked by the inner white curve and in the thick loop below it. Note that in the beginning of this time interval the loops overlap with each other, splitting into two distinct loops about one hour later. Of interest is the oscillation of the latter loop, indicated by the red curve in panel~(d). This oscillation lasts for about 3\,h with amplitude of 0.38\,Mm.
Figure~\ref{fig:im}(e) which shows another time interval, depicts transverse oscillatory patterns too. 
The hosting loop is highlighted by the dashed white curve in panel (b). The decayless oscillation of interest occurs at 10–14 h, see the red curve. This 4-hour oscillation sustains for more than 7 cycles with the oscillation period of a few tens of minutes and the displacement amplitude of 0.46\,Mm.

\begin{figure}[ht]
\centering
\includegraphics[width=\linewidth]{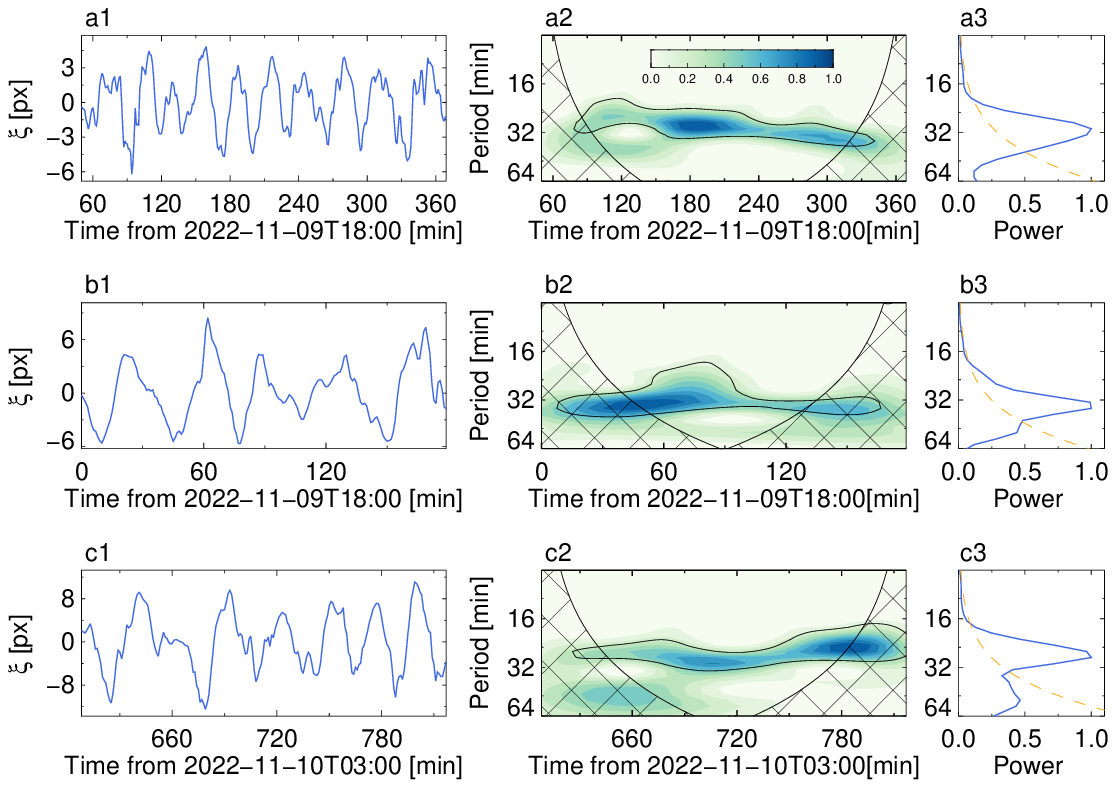}
\caption{Periodicities of the observed decayless kink oscillations. (a1--c1): Time series of oscillatory displacements extracted from the time--distance maps shown in Figure~\ref{fig:im}(c--e), with the background trend subtracted. (a2--c2): Wavelet power spectra normalised to their maxima. (a3--b3): Global wavelet power spectra. The black thick contour is the 95\% confidence level. The orange curves indicate the global 95\% confidence level. The filled hatch is the cone of influence.}
\label{fig:wavlet}
\end{figure}

Evolution of oscillation periods of the oscillations highlighted by the red curves in Figure~\ref{fig:im}(c--e), was studied with wavelet analysis. 
In Figure~\ref{fig:wavlet}, the columns from left to right display the detrended displacement signals, their wavelet spectra, and global wavelet spectra of three oscillations of interest. 
For all three events, most of the power is concentrated around the 30\,min period. The spectral peak associated with this periodicity has almost the same power during the event, indicating that the oscillation is apparently stationary during those time intervals.
Global wavelet spectra of these three oscillatory events peaks at 30.3\,min, 33\,min, and 28\,min respectively.


\subsection*{Thermal properties}
\label{sec:dem}
\begin{figure}[ht]
\centering
\includegraphics[width=\linewidth]{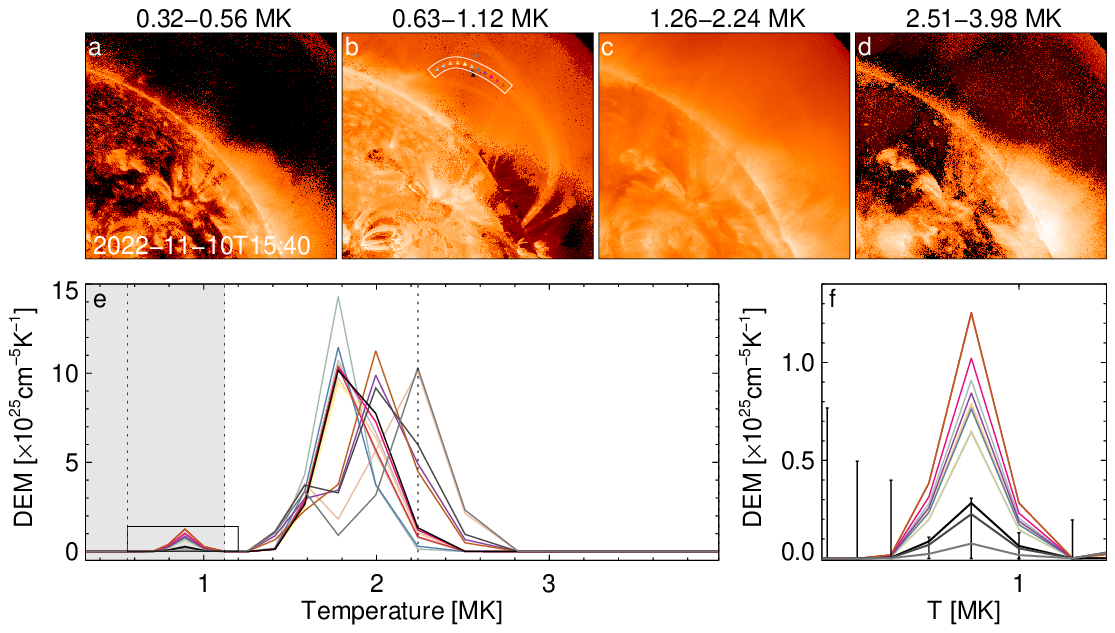}
\caption{Thermal properties of an oscillating loop at 2022-11-10T15:40 UT. (a--d): EM maps for different temperature ranges. (e--f): DEM distribution as a function of temperature for the selected locations (marked by the coloured triangles) inside and outside the analysed loop. The white slice highlights the loop segment which displays decayless oscillations, and this segment is used for plasma diagnostics. Panel (f) is a zoom-in version of the small window in (e). The vertical dotted lines divide the temperature range into 4 subsets as shown in panel (a--d). The grey area indicates the temperature range of the DEM integration for internal EM. The error bar is the 90\% confidence interval. }
\label{fig:DEM}
\end{figure}

Thermal properties of the plasma inside the loop which oscillates at around 2022-11-10T15:40 were established by the differential emission measure (DEM) technique with the use of AIA images taken at different EUV wavelengths \cite{2015ApJ...807..143C,2018ApJ...856L..17S}.
Figure \ref{fig:DEM} maps the emission measure (EM) of the thermal plasma in the region of interest from 0.3\,MK to 4\,MK, which is the typical temperature range for coronal loop. 
The loop plasma is most pronounced in the range of 0.63--1.12\,MK, and it is also partially seen in 1.26--2.24\,MK.

The DEM distributions as a function of temperature, are determined at 14 selected locations marked by the triangles filled in with different colours in Figure \ref{fig:DEM}(e). Among them there are 11 locations inside the loop, which coincide with the midpoints of the slits shown in Figure \ref{fig:im}b. Three other locations, indicated by black and gray triangles, are outside the loop, and give us information about the background emission.
DEM profiles of the internal and external points have two peaks, with the major peak at around 2\,MK, and the minor one at around 0.9\,MK. As shown in panel(c), for the major peak, the background peaks have the magnitude virtually equal to those inside the loop. It indicates that the contribution of EM in the 1.26--2.24\,MK range is mainly from the background plasma. Zooming in the range of 0.32--1.2\,MK, as displayed in panel (f), the background curves have much less magnitude in this minor peak.
Thus, we consider the temperature of the loop plasma to be mainly distributed in the range of 0.32--1.12\,MK. 

\begin{figure}[ht]
\centering
\includegraphics[width=\linewidth]{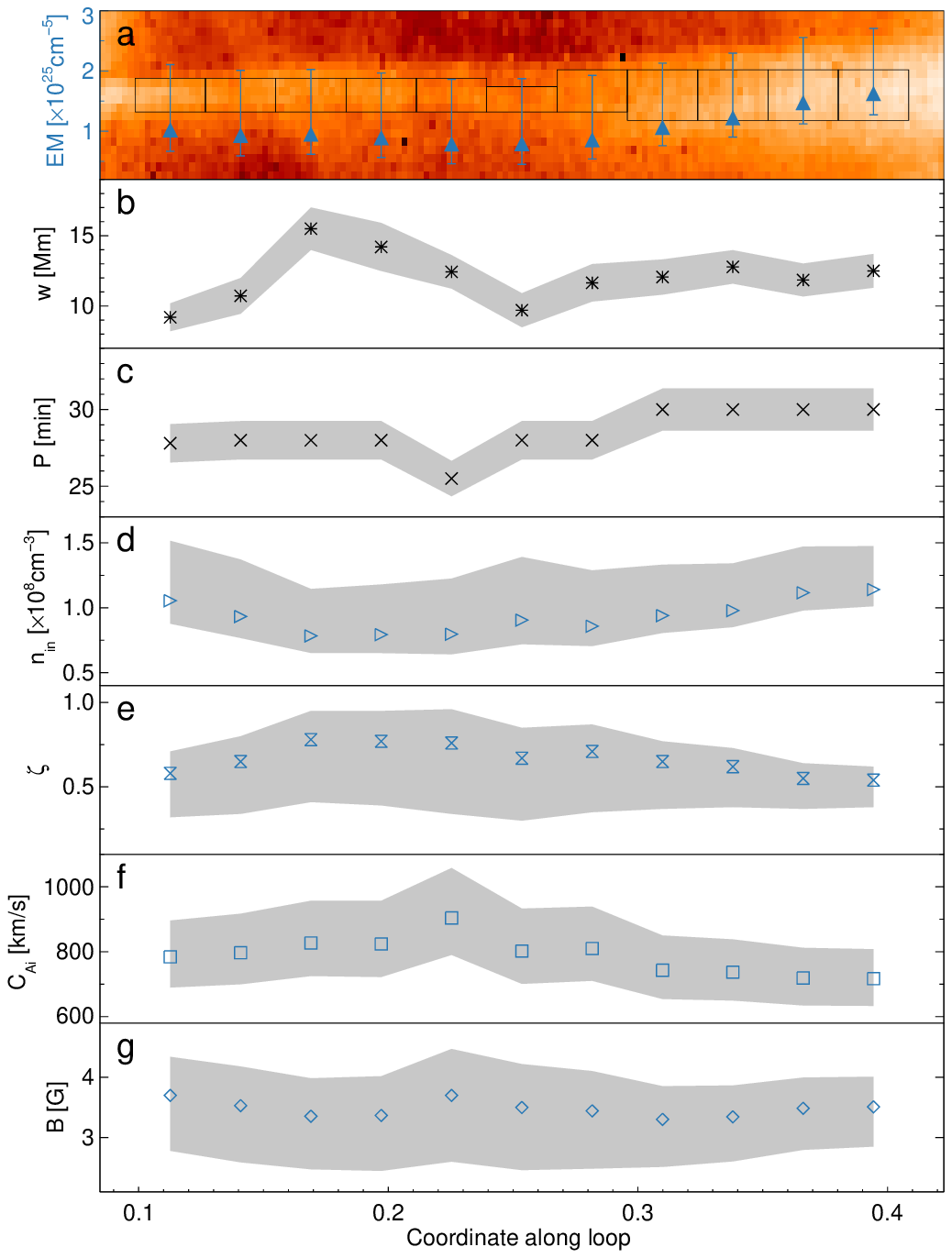}
\caption{Plasma diagnostics in the oscillating loop. (a): EM maps over 0.32--1.12\,MK of the loop segment marked by the white slice in Figure~\ref{fig:DEM}. The blue triangles indicate the EM averaged over the area marked by the black rectangles. (b--g): the variation of various parameters along the selected segment: the loop width $w$ (FWHM) (b), oscillation period $P$ (c), electron density inside the loop $n_{in}$ (d), density contrast of external and internal density $\zeta$ (e), internal Alfv\'en speed $C_{Ai}$ (f), and the magnetic field strength $B$ (g). 
 The filled polygons indicate the error bars.}
\label{fig:seis}
\end{figure}

Figure \ref{fig:seis}(a) features the EM map from 0.32\,MK to 1.12\,MK of the loop segment where decayless oscillations are resolved, as marked by the white slice in Figure \ref{fig:DEM}(b). For each position indexed by the slit number or the coordinate along the loop, the EM is averaged over the area indicated by the corresponding black rectangle, after subtracting the average background distribution.  
From slit 8 to 28, EM ranges from 0.78$\times 10^{25}$\,cm$^{-5}$ to 1.61$\times 10^{25}$\,cm$^{-5}$, with the average value of 1.04$\times 10^{25}$\,cm$^{-5}$, and the standard derivation of 0.27$\times 10^{25}$\,cm$^{-5}$, see more details in Table \ref{tab:example}. Besides, the DEM-weighted mean temperature of this coronal loop is 0.88\,MK, and that of background corona is 1.84\,MK. 

Inferred from the background EM (see Methods), the external number density of loop is estimated as $(6.09\pm0.55)\times 10 ^{7}$\,cm$^{-3}$.
Likewise, the electron density inside the loop $n_{in}$ is derived to span from 0.78 $\times 10^{8}\,\mathrm{cm}^{-3}$ to 1.14$\times 10^{8}$\,cm$^{-3}$, with the average of 0.9$^{+0.4}_{-0.2}\times 10^{8}\,\mathrm{cm}^{-3}$, see Figure \ref{fig:seis}(d). This estimation is consistent with other studies \cite{2011ApJ...736..102A,2014LRSP...11....4R}. 
There have also been estimations showing higher densities, e.g., about $10^{9}$\,cm$^{-3}$ \cite{2005ApJ...633..499A,2013SoPh..283....5A,2016NatPh..12..179J,2020A&A...639A.114L}. But, those estimations are made at lower heights in comparison of our estimation made at the projected height of 234\,Mm. 
 In that case, the discrepancy could be attributed to the effect of gravitational stratification.

The ratio of the number densities outside and inside the loop, $\zeta = n_\mathrm{ex}/n_\mathrm{in}$, is the key parameter in the theory of kink oscillations \cite{1983SoPh...88..179E}.
As shown in Figure \ref{fig:seis}(e), the density contrast of 0.54–-0.78 with the error up to 0.42 is consistent by the order of magnitude with the characteristic range of 0.07-–0.66 shown in \cite{2003ApJ...598.1375A} and 0.78--0.87 obtained in \cite{2008A&A...487L..17V}. 


\subsection*{Seismology}
\label{sec:seismology}

In the context of coronal seismology, observables in the detected kink oscillation are used to estimate the kink speed $C_k$, Alfv\'en speed $C_{Ai}$, and the absolute value of the magnetic field $B$, see details in Methods.
Figure \ref{fig:seis}(b--g) shows parameters of the oscillation and properties of the loop along a selected loop segment. Additional details are given in Table \ref{tab:example}.
 
As displayed in Figure \ref{fig:seis}b, the loop width is around 12\,Mm.
In panel (d), oscillation periods vary from 25.5\,min to 30\,min along the selected loop segment.
The estimated internal Alfv\'en speed ranges from 717 to 904\,km/s, which agrees with previous observations \cite{2019ApJ...884L..40A}.
The magnetic field strength of 3.3--3.7\,G is consistent with previous seismological inversions as well.

As shown in Table \ref{tab:example}, parameters of the oscillations and properties of the loop are nearly homogeneous  along the loop, with minor variance. After averaging, we obtain the average loop width of $12.1\pm1.3$\,Mm, the oscillation period of $28.5\pm1.3$\,min, the kind speed of $864\pm98$\,km/s, the internal density of $0.9^{+0.4}_{-0.2}\times10^{8}$\,cm$^{-3}$,
the density contrast of $0.7^{+0.1}_{-0.3}$, 
the internal Alfv\'en speed of $788^{+118}_{-96}$\,km/s, 
and the magnetic field strength of $3.5^{+0.6}_{-0.9}$\,G. 
In addition, the average external Alfv\'en speed is $971^{+377}_{-144}$\,km/s. 
This agrees with a previous seismological study of a trans-equatorial loop system which exhibited a decaying oscillation in the second harmonic with the period of 18\,min, with loop length of $711\pm11$\,Mm, internal density of $(1.1\pm 0.5) \times 10^8$\,cm$^{-3}$, external density of $(2.5\pm 2.1) \times 10^7$\,cm$^{-3}$, magnetic field strength of $5.5\pm1.5$\,G \cite{2017A&A...603A.101L}. This previous result was further supported by other two approaches based on the use of a transverse wave propagation and magnetic field extrapolation.
Another example of similar diagnostics by a decaying kink oscillation of a shorter loop with length of $143\pm20$\,Mm yielded $P=6.26$\,min, $n_{in}=(1.9\pm0.3)\times10^8\,\mathrm{cm}^{-3}$, $w=4.9\pm0.6$\,Mm, $C_{Ai}=560\pm100$\,km/s, B$=4.0\pm0.7$\,G \cite{2011ApJ...736..102A}. 
{}{On the other hand, in a dipolar magnetic field, the magnetic field strength decreases with height proportionally to the ratio of height and solar radii to the power of 3, $B_r=B_0(r/R_{sun})^{3}$. Assuming the strength at the footpoint $B_0$ is 100\,G, the value at the loop top (234\,Mm) is about 3.8\,G, which is commensurate to our estimations.}
Hence we consider the measurements made in the current study to be trustworthy. 

\section*{Discussion}
\label{sec:discussion}

\begin{figure}[ht]
\centering
\includegraphics[width=\linewidth]{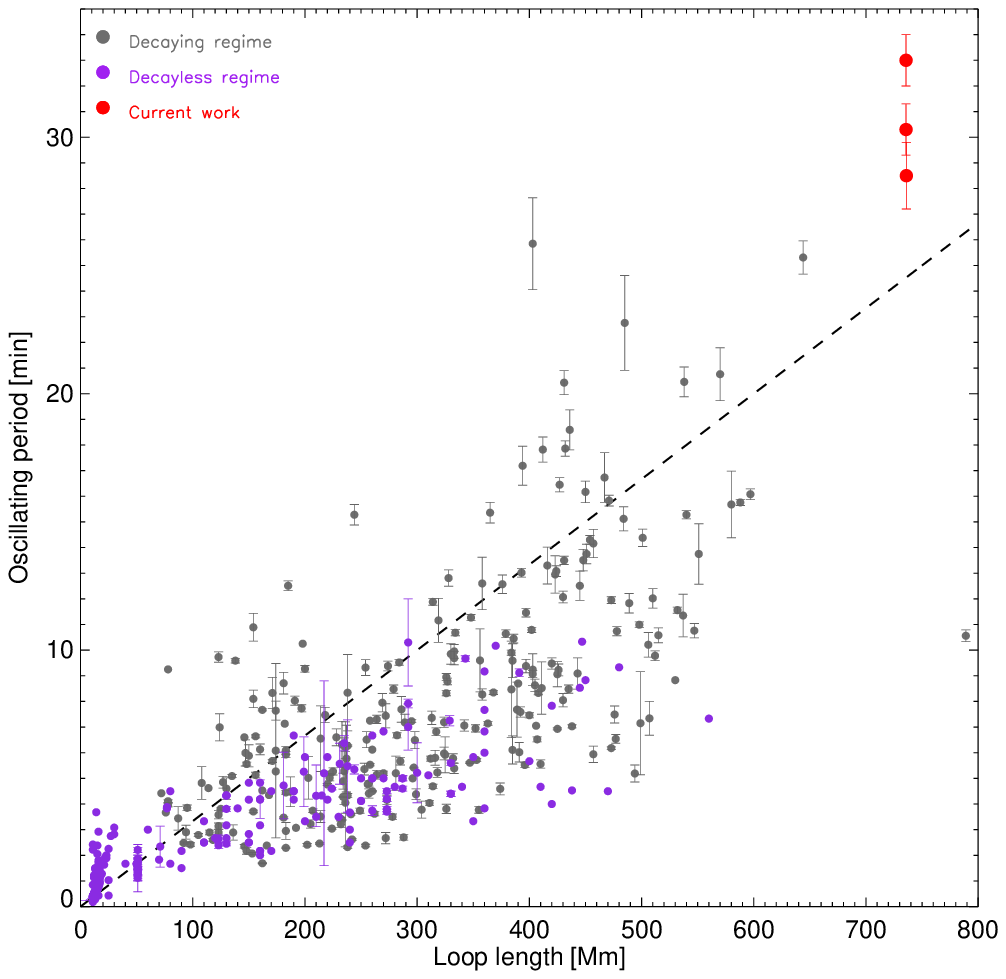}
\caption{Scaling of loop length and oscillation period in observed kink oscillations in both decaying and decayless regimes. For the decaying cases (in gray), 252 events are collected from \cite{1999ApJ...520..880A,2002SoPh..206...99A,2003ApJ...598.1375A,2004SoPh..223...77V,2013ApJ...777...17S,2019ApJS..241...31N,2021A&A...652L...3M,2022RAA....22k5012G}.
For the decayless cases (in purple), parameters of \textbf{238} oscillations are extracted from \cite{2012ApJ...751L..27W,2013A&A...552A..57N,2015A&A...583A.136A,2019ApJ...884L..40A,2018ApJ...854L...5D,2021A&A...652L...3M,2022MNRAS.513.1834Z,2022MNRAS.516.5989Z,2022A&A...666L...2M,2023ApJ...946...36P,2023ApJ...944....8L}.
The dashed line represents the kink speed of 1000\,km/s.
}
\label{fig:scale}
\end{figure}

In this work, very-long coronal loops forming a long-living bundle are found to exhibit decayless kink oscillations with periods of  28--33\,min and amplitudes of 0.3--0.5 Mm. The observed oscillation amplitudes are similar to those reported in events with shorter periods.
{}{The main findings, interpretations, and conclusions are as follows.}
\begin{enumerate}
\item This 30-min periodicity exceeds the decayless kink oscillation period ever reported \cite{2019ApJS..241...31N,2015A&A...583A.136A}, both in decaying and decayless regimes, see Figure \ref{fig:scale}. The length of one of the oscillating loops is $736\pm80$\,Mm, is about 30\% longer than the longest loop performing any previously detection decayless kink oscillations.
The estimated average kink speed of $788^{+118}_{-96}$\,km/s 
is within the range of previously detected kink oscillations \cite{2015A&A...577A...4Z,2016A&A...585A.137G,2019ApJS..241...31N}. 
This finding extends the range in the scaling of the loop length and decayless kink oscillation period, and further justifies that a certain mechanism sustaining the coronal energy balance effective works throughout a broad range of oscillation periods and loop lengths.

\item Such long periodicity of a fundamental mode departs away by an order of magnitude from the 3-min chromospheric oscillations or 5-min p-modes which have been often considered as the drivers of coronal waves and oscillations. Thus, our result provides an additional evidence for the exclusion of resonant excitation mechanisms for decayless kink oscillations \cite{2016A&A...591L...5N}. 
Recall that in the self-oscillatory or random driving model, the oscillation period is independent of the driver \cite{2020ApJ...897L..35K,2021ApJ...908L...7K}.
So both these mechanisms are able to sustain the 30-min decayless kink oscillations. In both those cases the driver utilises the energy in the low-frequency part of the spectrum, indicating its transfer to the corona.


\item Considering that the oscillating loop has length of $736$\,Mm, diameter of 12\,Mm, number density of $9\times10^{7}$\,cm$^{-3}$, oscillation period of 30\,min, and amplitude of around 0.4 Mm, the kinetic energy of the observed decayless kink oscillation is about $1.2\times10^{23}$\,erg per cycle, 
which is comparable to the energy of a nanoflare ($10^{23}$--$10^{25}$\,erg). This value is 3 order of magnitude smaller than that of a typical decaying kink oscillation, up to $10^{26}$\,erg \cite{2007A&A...469.1135T,2020ARA&A..58..441N}. 
Note that LoS projection effect decreases the apparent displacement amplitude in the plane of the sky, hence the kinetic energy is somehow underestimated.
The total energy converted into heat by this process could be far higher than the estimated kinetic energy of the oscillation, as the amplitude is not a proxy of the energy budget and their relationship is unsettled\cite{2019FrASS...6...38K}. To be more specific, the oscillation amplitude is a indicator of the reminder of the energy input and dissipation. 
From the perspective of the energy balance, at least $10^{23}$\,erg per oscillation cycle is needed to be supplied continuously, for several hours at least, to sustain the observed oscillations. The ubiquity of the decayless kink oscillations suggests that such an energy supply is ubiquitous too. The energy transport and conversion in such a steady fashion, possibly via the interactions between loops and steady{}{/random} flows, 
demonstrates the efficiency and potential to sustain the million Kelvin temperature of the corona.
{An interesting question is the reason for the external plasma being hotter than the internal plasma in the oscillating loop, about 1.8 MK versus 0.9 MK according to our estimations. It may be connected with lower radiative losses in the more rarefied external plasma, as well as with the  dependence of the optically thin radiation upon the temperature. The heating of the external medium could be carried out by the evanescent or leaky part of the oscillation. A detailed study of this phenomenon would be of interest in the future.}  




\item The seismological analysis performed in our study, supplemented with the DEM analysis, returns reliable estimation of the loop properties, such as the electron density, density contrast, Alfv\'en speed, and, particularly, the magnetic field strength of $3.5^{+0.6}_{-0.9}$\,G.
The magnetic field strength is crucial for the attempting magnetic energy and tracing the energy flow. 
Such detailed measurement of physical parameters in a coronal loop showing decayless kink oscillations with reliably determined parameters, can, in return, impose important constrains on the mechanism sustaining decayless kink oscillations. This therefore calls for further theoretical works.

\end{enumerate}

\section*{Methods}
\label{sec:method}


\subsection*{Data processing and oscillation analysis}
\label{sec:method_obs}
The oscillating loops are studied by analysing time sequences of EUV images obtained with AIA from 2022-11-08T12:00 to 2022-11-10T20:30 UT, with the pixel size of 0.6\,arcsec and time cadence of 1\,min. 
The AIA level 1 data obtained in six EUV passbands (94, 131, 171, 193, 211, and 335\,\AA) {are} downloaded form the JSOC center.
The image sequences are processed using the SSWIDL routine \texttt{aia\_prep.pro}. Since the loops of interest are off the solar limb, there is no need for de-rotation.
In the oscillation analysis, we mainly use 171\,\AA\ images, which provide us with the highest contrast.

Given that the characteristic amplitude of decayless kink oscillations is less than the pixel size of AIA images, the motion magnification technique is implemented to amplify the tiny transverse quasi-periodic displacements. This technique is based on the dual-tree complex wavelet transform-based motion magnification algorithm\cite{2016SoPh..291.3251A}, which has been successfully applied to analyse decayless kink oscillations \cite{2018ApJ...854L...5D, 2019ApJ...884L..40A,2021A&A...652L...3M, 2022ApJ...930...55G, 2022MNRAS.516.5989Z}. 
This algorithm first decomposes the input image sequence into images with different scales via wavelet transform, and for each pixel, their phases are calculated. The variation of phase depends on the motion in the plane of the sky, which is the key feature of dual-tree complex wavelet transform.
This algorithm takes two parameters set by the user: the smoothing width that is used to obtain the phase trend by smoothing, and the magnification factor that are used in linear magnification by multiplying the relative phase.
Finally, the magnified image sequence is reconstructed using the inverse dual-tree complex wavelet transform.
In this work, the 171\,\AA\ image sequence are magnified with a magnification factor of 5 and smoothing width slightly longer than the period.

In addition to visual inspections, time--distance maps are typically useful to detect and analyse oscillations.
Slits are put across the loop. 
The slits are sufficiently long to include the whole loop width.
Each slit has width of 5 pixels. For each time frame, the intensity along the slit is averaged over the width to reduce the uncertainty. 
Several such slits are evenly constructed along the loop. 
Extracted from the time-distance map, the oscillation is traced by the instantaneous best fitting transverse intensity profile of a Gaussian shape, at each instance of time with the use of the \texttt{GaussFit.pro} routine.
For each Gaussian fit, we obtain the Gaussian centre and full width at half maximum (FWHM), as well as their uncertainties, which corresponds to the location of the loop centre and loop width (minor diameter) respectively.
In this work, we use the mean loop width which is the average of instantaneous widths over the lifetime of oscillation.

The oscillation signals are detrended by subtracting the trends which are obtained by smoothing the original with a window slightly longer than the expected period (40\,min).
For each time series (detrended), we employ the Hilbert transform to calculate their instantaneous amplitudes.

\subsection*{Wavelet analysis}

Wavelet analysis \cite{1998BAMS...79...61T} is performed to estimate the periodicity and its variance in the time series data, using the Morlet mother function.
The global wavelet spectrum is the averaged local wavelet spectrum over time span. 
Errors in the estimation of the oscillation period are determined by the size of the half-bin.
The wavelet software was provided by C. Torrence and G. Compo, and is available at: \url{http://paos.colorado.edu/research/wavelets/}.

\subsection*{DEM analysis}
\label{sec:method_dem}
Plasma differential emission measures (DEMs) are derived from AIA EUV images in six channels (94, 131, 171, 193, 211, and 335\,\AA), using the sparse inversion code \cite{2015ApJ...807..143C} with the updated solution \cite{2018ApJ...856L..17S}. The latest version can be found in \cite{2022ApJ...930..147L}.
In this work, we take images at around 2022-11-10T15:40 with the same FOV as shown in Figure \ref{fig:im}b for DEM analysis. The images are rebinned by $3\times3$\,pixels to reduce the noise.
As a result, we get the spatial distribution DEM as a function of temperature ranging from 0.3 to 30\,MK. 
The uncertainty of DEM is estimated by 100 repetitive analyses with the contamination by noise via a Monte Carlo simulation \cite{2018ApJ...856L..17S}. Here we take the 90\% credible bounds as errors. 
The emission measure (EM) is calculated as
\begin{equation}
    \textrm{EM} = \int \textrm{DEM}(T)\,dT,
\end{equation}
where $T$ is the temperature.
The DEM-weighted mean temperature $\overline{T}$ is estimated by \cite{2012ApJ...761...62C} 
\begin{equation}
    \overline{T}=\frac{\int \textrm{DEM}(T)\,T\,dT}{\int \textrm{DEM}(T)\,dT}\\
\end{equation}

Given that EM is the line-of-sight (LoS) integration of the density squared, the electron density outside the analysed loop $n_\mathrm{ex}$ is calculated as:
\begin{equation}
\label{eq:ex_density}
    n_\mathrm{ex} = \sqrt{\mathrm{EM}_\mathrm{bg}/s},
\end{equation}
where $\mathrm{EM}_{bg}$ is averaged over a region right below the loop, and $s$ is the LoS path length, which is inferred from the heliocentric distance $r$ and scale height $\Lambda$, according to the geometrical model of a stellar atmosphere by \cite{2014A&A...564A..47Z,1936HarCi.417....1M}, as 
\begin{equation}
\label{eq:los_path}
    s=\sqrt{\Lambda \pi r}\\
\end{equation}
where $\Lambda=84$\,Mm at the temperature of the background corona $\overline{T}=1.8$\,MK, and $r=930$\,Mm is the sum of the solar radius, 696\,Mm, and the major radius of the loop, 234\,Mm, giving $s=495.4$\,Mm.

To estimate the electron density inside the loop, we compare the DEM distributions in the loop and background to extract the EM inside the loop, $\mathrm{EM}_\mathrm{in}$. 
As inferred from Figure \ref{fig:DEM}, the loop plasma is mainly distributed in the range of 0.3--1.12\,MK, so $\mathrm{EM}_\mathrm{in}$ is DEM integration in this temperature range after the subtraction of the background EM. The background EM is the average of three background locations (the black and gray triangles in Figure~\ref{fig:DEM}) in the same temperature range.
Also, the LoS column depth for $\mathrm{EM}_\mathrm{in}$ is loop width $w$, therefore the electron density inside the loop $n_\mathrm{in}$ is calculated by Eq.~(\ref{eq:density}) in \cite{2011ApJ...736..102A}, assuming the filling factor of unity. 
\begin{equation}
\label{eq:density}
    n_\mathrm{in} = \sqrt{\mathrm{EM}_\mathrm{in}/w}\\
\end{equation}

\subsection*{MHD wave model}

Waves and oscillations in a coronal loop are modelled as linear magnetohydradynamic (MHD) waves in a thin magnetic cylinder by Zajtsev \& Stepanov \cite{1975IGAFS..37....3Z} and Edwin \& Roberts \cite{1983SoPh...88..179E}. The magnetic field is considered to be straight and directed along the axis the cylinder. In the long wavelength limit, kink waves with the azimuthal wave numbers $m=1$ have the phase speed approaching the so-called kink speed,
\begin{equation*}
    C_{k} = (\frac{\rho_{0i}{C_{Ai}}^{2}+\rho_{0e}{C_{Ae}}^{2}}{\rho_{0i}+\rho_{0e}})^{1/2}\\
\end{equation*}
where $\rho_{0i}$ and $\rho_{0e} $ is the internal (index by $i$) and external (index by $e$) mass density of the cylinder respectively, $C_{Ai}$ and $C_{Ae}$ is the internal and external Alfv\'en speed respectively. In zero-$\beta$ limit, where the magnetic fields inside and outside the cylinder are identical, the kink speed reduces to
\begin{equation}\label{Eq:alfven}
    C_\mathrm{k} \approx C_{Ai} \sqrt{\frac{2}{1+\zeta}}\\
\end{equation}
where $\zeta=\rho_{0e}/\rho_{0i} = n_\mathrm{ex}/n_\mathrm{in}$ is the density contrast of the cylinder.

MHD seismology is a technique combining MHD wave theory and observed parameters of the observed wave to infer the physical condition of plasma \cite{1984ApJ...279..857R}. 
For fundamental standing kink mode, neglecting the dispersive effects, the phase speed depends on the ratio of the loop length $L$ and oscillation period $P$:
\begin{equation}
    C_\mathrm{k} = 2L/P.
\end{equation}
Neglecting effects of the varying cross-section and stratification, in the low-$\beta$ limit, we can estimate the internal Alfv\'en speed $C_{Ai}$ with the use of Equation~(\ref{Eq:alfven}),
with the internal number density inferred from EM, see Equation~(\ref{eq:density}), and the external density of the surrounding corona $n_\mathrm{ex}$ is calculated by Equation~(\ref{eq:ex_density}).
The magnetic filed strength $B$ can be estimated as  
\begin{equation}
    B =C_{Ai}\sqrt{\mu_{0}\rho_{0i}},
\end{equation}
where $\mu_0$ is the magnetic permeability (\cite{1984ApJ...279..857R,2001A&A...372L..53N,2011ApJ...736..102A}).

For the 11 slits along the loop, we first extract the time series of oscillation from the corresponding time--distance map, and calculate the oscillation period (Figure~\ref{fig:seis}d) via wavelet analysis, estimate the number density and density contrast from DEM analysis, and then use those values in the seismological inversions resulting in $C_\mathrm{k}$, $C_{Ai}$ (Figure \ref{fig:seis}f), B (Figure \ref{fig:seis}g) along the loop segment. 
The uncertainties of the derived parameters above are estimated by the error propagation formula \cite{error2010}, based on the errors of those independent variables including $L$, $P$, EM$_\mathrm{bg}$, EM$_\mathrm{in}$, and $w$.

\section*{Data availability}
The data analysed during the current study are obtained from the Joint Science Operations Center (JSOC) database (\url{http://jsoc.stanford.edu/}).




\section*{Acknowledgements}

S.Z. acknowledges support from China Scholarship Council-University of Warwick joint scholarship.
Y.M. is supported by the National Natural Science Foundation of China (NSFC, 12102016), the Fund of Shenzhen Institute of Information Technology (SZIIT2022KJ040), project of Shenzhen Science and Technology Innovation Committee (JSGG20211029095003004).
D.Y. and L.F. are supported by the National Natural Science Foundation of China (NSFC,12173012,12111530078), the Guangdong Natural Science Funds for Distinguished Young Scholar (2023B1515020049), the Shenzhen Technology Project (GXWD20201230155427003-20200804151658001), and the Shenzhen Key Laboratory Launching Project (No. ZDSYS20210702140800001).

\section*{Author contributions statement}


V.M.N. discovered the event and conceived the problem. 
S.Z. performed the event analysis including data processing, oscillation analysis, DEM analysis, and plasma diagnostics, and wrote the first draft of the manuscript. L.F. performed the wavelet analysis. Y.M. has contributed to the DEM analysis. V.M.N. and D.Y. provide guidelines for scientific research and supervised the completion of the project. All authors reviewed the manuscript.

\section*{Competing interests statement}
The authors declare that they have no conflicts of interests.



\setlength{\extrarowheight}{8pt}
\begin{table}[ht]
\centering
\begin{tabular}{ccccccccc}
\hline
Slit number & $w$ [Mm] & $P$ [min] & $C_k $[km/s] & EM$_{in}$ [$\times 10^{25}\mathrm{cm}^{-5}$] & $n_{in} [\times 10^{8} \mathrm{cm}^{-3}]$ & $\zeta$ & $C_{Ai} $[km/s] & B [G]\\
\hline
8 & $9.2\pm 1.0$ & $27.8\pm1.25$ & $883\pm 100$ & $1.01^{+1.38}_{-0.35}$& $1.05^{+0.46}_{-0.18}$ & $0.58^{+0.13}_{-0.26}$  & $784^{+112}_{-94}$ & $3.7^{+0.6}_{-0.9}$ \\
\hline
10 & $10.7\pm 1.3$ & $28\pm1.25$ & $877\pm 100$ & $0.92^{+1.37}_{-0.33}$& $0.93^{+0.44}_{-0.17}$ & $0.65^{+0.15}_{-0.31}$ & $797^{+120}_{-95}$ & $3.5^{+0.7}_{-0.9}$ \\
\hline
12 & $15.5\pm 1.5$ & $28\pm1.25$ & $877\pm 100$ & $0.94^{+1.37}_{-0.32}$& $0.78^{+0.36}_{-0.13}$ & $0.78^{+0.17}_{-0.38}$ & $827^{+130}_{-102}$ & $3.4^{+0.6}_{-0.9}$ \\
\hline
14 & $14.2\pm 1.7$ & $28\pm1.25$ & $877\pm 100$ & $0.88^{+1.37}_{-0.32}$& $0.79^{+0.39}_{-0.14}$ & $0.77^{+0.18}_{-0.38}$ & $824^{+133}_{-102}$ & $3.4^{+0.7}_{-0.9}$\\
\hline
16 & $12.4\pm 1.2$ & $25.5\pm1.15$ & $962\pm 110$ & $0.78^{+1.37}_{-0.32}$& $0.80^{+0.43}_{-0.16}$ & $0.76^{+0.20}_{-0.42}$ & $904^{+154}_{-114}$ & $3.7^{+0.8}_{-1.1}$ \\
\hline
18 & $9.7\pm 1.2$ & $28\pm1.25$ & $877\pm 100$ & $0.78^{+1.37}_{-0.33}$& $0.90^{+0.49}_{-0.19}$ & $0.67^{+0.18}_{-0.37}$ & $802^{+131}_{-101}$ & $3.5^{+0.7}_{-1.0}$ \\
\hline
20 & $11.7\pm 1.3$ & $28\pm1.25$ & $877\pm 100$ & $0.84^{+1.37}_{-0.30}$& $0.86^{+0.43}_{-0.15}$ & $0.71^{+0.16}_{-0.36}$ & $810^{+129}_{-100}$ & $3.4^{+0.7}_{-1.0}$ \\
\hline
22 & $12.1\pm 1.3$ & $30\pm1.37$ & $818\pm 93$ & $1.05^{+1.37}_{-0.29}$& $0.94^{+0.39}_{-0.13}$ & $0.65^{+0.12}_{-0.28}$ & $743^{+107}_{-89}$ & $3.3^{+0.6}_{-0.8}$ \\
\hline
24 & $12.8\pm 1.2$ & $30\pm1.37$ & $818\pm 93$ & $1.21^{+1.37}_{-0.31}$& $0.98^{+0.36}_{-0.13}$ & $0.62^{+0.11}_{-0.24}$ & $737^{+101}_{-88}$ & $3.3^{+0.5}_{-0.7}$ \\
\hline
26 & $11.9\pm 1.2$ & $30\pm1.37$ & $818\pm 93$ & $1.46^{+1.38}_{-0.34}$& $1.12^{+0.36}_{-0.14}$ & $0.55^{+0.09}_{-0.18}$ & $719^{+93}_{-85}$ & $3.5^{+0.5}_{-0.7}$ \\
\hline
28 & $12.5\pm 1.2$ & $30\pm1.37$ & $818\pm 93$ & $1.61^{+1.38}_{-0.34}$& $1.14^{+0.33}_{-0.13}$ & $0.54^{+0.08}_{-0.16}$ & $717^{+91}_{-84}$ & $3.5^{+0.5}_{-0.7}$ \\
\hline
Average & $12.1\pm 1.3$ & $28.5\pm1.3$ & $864\pm 98$ & $1.04^{+1.37}_{-0.32}$& $0.94^{+0.40}_{-0.15}$ & $0.66^{+0.14}_{-0.30}$ & $788^{+118}_{-96}$ & $3.5^{+0.6}_{-0.9}$ \\
\hline
\end{tabular}
\caption{\label{tab:example}Physical properties inside the analysed loop segment, including the loop width $w$, oscillation period $P$, phase speed $C_k$, background-subtracted EM inside the loop EM$_{in}$, internal number density $n_{in}$, density contrast $\zeta$, internal Alfv\'en speed $C_{Ai}$, and the magnetic field strength $B$. 
}
\end{table}

\end{document}